\documentclass[10pt,twocolumn,floatfix,prb,aps,showpacs,superscriptaddress,longbibliography]{revtex4-2}
\usepackage{graphicx,amsmath,amssymb,color, siunitx, gensymb}
\usepackage{multirow}
\usepackage{nicefrac}
\usepackage[titletoc,title]{appendix}
\usepackage{amsmath}
\usepackage{subfigure}
\usepackage{tabularx}
\usepackage{xcolor}
\usepackage{tabularx}
\usepackage{booktabs}
\usepackage[colorlinks,bookmarks=true,citecolor=blue,linkcolor=red,urlcolor=blue]{hyperref}
\usepackage{titlesec}

\makeatletter
\def\@fnsymbol#1{%
  \ifcase#1\relax 
  \or \ensuremath{\color{blue}\dagger}
  \or \ensuremath{\color{red}*}
  \or \ensuremath{\mathsection}
  \or \ensuremath{\mathparagraph}
  \else\@ctrerr\fi
}
\makeatother

\begin{document}
	
	\title{Magnetic properties of a buckled honeycomb lattice antiferromagnet}
    
    \author{A. Yadav}
	\thanks{equal contribution}
    \affiliation{Department of Physics, Indian Institute of Technology Madras, Chennai, 600036, India}
    \author{U. Jena}
    \thanks{equal contribution}
	\affiliation{Department of Physics, Indian Institute of Technology Madras, Chennai, 600036, India}

    \author{A. Pradhan}
	\affiliation{Department of Physics, Indian Institute of Technology Madras, Chennai, 600036, India}
    \author{Satish K.}
	\affiliation{Department of Physics, Indian Institute of Technology Madras, Chennai, 600036, India}
   
\author{P. Khuntia}
\email{pkhuntia@iitm.ac.in}
\affiliation{Department of Physics, Indian Institute of Technology Madras, Chennai, 600036, India}
\affiliation{Quantum Centre of Excellence for Diamond and Emergent Materials,
Indian Institute of Technology Madras, Chennai, 600036, India}

\date{\today}
\begin{abstract} 
The intriguing interplay between competing degrees of freedom in frustrated magnets can lead to non-trivial magnetic phenomena with exotic low-energy excitations that are highly relevant for addressing some of the fundamental questions in quantum condensed matter as well as potential technological applications. Herein, we report the synthesis and thermodynamic results on a frustrated magnet Co$_3$ZnNb$_2$O$_9$. The Co$^{2+}$ moments constitute buckled AB-type honeycomb layers in the $ab$-plane. The temperature-dependent magnetic susceptibility shows a sharp anomaly at $\sim14$~K, indicating the onset of long-range magnetic ordering. The Curie-Weiss fit of the magnetic susceptibility above $100$~K, yields a Curie-Weiss temperature of $-70$~K, suggesting strong antiferromagnetic (AFM) interactions between the Co$^{2+}$ spins and an effective magnetic moment of 5.54 $\mu_B$, indicating the presence of unquenched orbital angular momentum. A field-induced spin-flop–like metamagnetic transition below the ordering temperature is characterized by a critical magnetic field of $\sim1.2$ T. The specific heat shows a $\lambda$-type anomaly at 14 K, confirming the presence of long-range magnetic ordering, due to finite interlayer interaction. Interestingly, our study of the magnetocaloric effect near the transition temperature revealed an entropy change of 2.81 J/kg·K, which is ascribed to competing interactions,  underlying anisotropy, and reduced net magnetization lead to relatively small isothermal entropy changes that suggest that frustrated  honeycomb magnets are promising contenders for field-induced exotic phases and magnetocaloric response. 

~~~~~~~~~~~~~~~~~~~~~~~

\end{abstract}
\maketitle

\section{Introduction} 

In strongly correlated quantum materials, the interplay between spin, charge, orbital, and lattice degrees of freedom gives rise to emergent collective phenomena that elude standard paradigms~\cite{laughlin2000theory,khatua2023experimental,witczak2014correlated,PhysRevB.106.104404,PhysRevLett.113.216403,PhysRevB.90.035141}. Among these, the coupling between magnetic and electric order parameters, often termed as  `magnetoelectric (ME) coupling', is particularly appealing as it links two fundamentally distinct symmetries, namely time reversal and spatial inversion within a single framework~\cite{landau2013electrodynamics, mostovoy2024multiferroics, hill2000there}. Multiferroics are functional materials that exhibit the coexistence of at least two primary ferroic orders—typically ferromagnetism, ferroelectricity, or ferroelasticity—within a single phase~\cite{khomskii2009classifying}. Specifically, type-II multiferroics represent a class of correlated magnets wherein ferroelectricity emerges as an emerging order parameter at the same temperature scale as magnetic ordering, with each mechanism driving and reinforcing the other~\cite{spaldin2005renaissance,fiebig2016evolution}.  
However, such multiferroics are rare, as magnetism requires partially filled $d$-orbitals, whereas ferroelectricity favors $d^0$ electronic configuration. Overcoming this apparent incompatibility has led to the search for unconventional mechanisms in strongly correlated magnets, where non-collinear or non-centrosymmetric spin textures—such as cycloidal, helical, or exchange-striction–driven collinear orders can induce electric polarization~\cite{mostovoy2024multiferroics}. Understanding and controlling such intertwined orders are fundamentally important and technologically promising, thereby motivating the search for new multifunctional materials exhibit strong magnetoelectric responses~\cite{eerenstein2006multiferroic}. Most type-II multiferroics emerge from magnetically frustrated materials, where  competing interactions stabilize non-collinear spin textures that break spatial inversion symmetry and induce ferroelectricity~\cite{cheong2007multiferroics,PhysRevLett.101.067204}. 

\begin{figure*}[ht]
\includegraphics[width=0.99\textwidth]{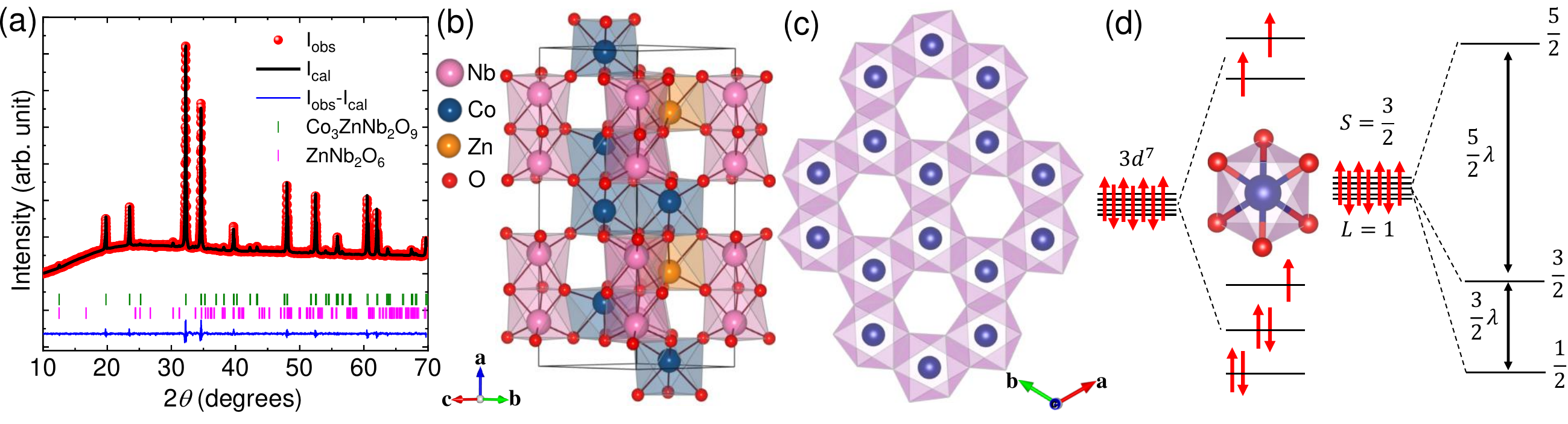}
  \caption{(a) Rietveld refinement of the powder X-ray diffraction data of Co$_3$ZnNb$_2$O$_9$ taken at room temperature. (b) Crystal structure of Co$_3$ZnNb$_2$O$_9$. (c) The buckled honeycomb lattice formed by the Co$^{2+}$ moments in the crystallographic $ab$ plane. (d) Splitting of energy levels for a high-spin $3d^7$ ion (Co$^{2+}$) in an edge-sharing octahedral environment.}
\label{CrystalStructure}
\end{figure*}

Magnetic frustration is the incompatibility of simultaneously satisfying all interactions in a magnetic material owing to geometrical constraints and competing exchange interactions, leading to a macroscopically degenerate ground-state manifold~\cite{PhysRev.102.1008}. The transition-metal-based frustrated magnets provide a fertile ground for the experimental realization of many emergent quantum phenomena owing to their localized $d$-orbitals, which enhance Coulomb interactions and reduce bandwidth, thereby magnifying the effect of magnetic frustration and competing interactions~\cite{ khuntia2019novel,witczak2014correlated}. Recently, considerable attention have been devoted to 4$d$/5$d$ transition metal-based honeycomb magnetic oxides due to their potential realization of exotic quantum phases including Kitaev spin liquid with non-trivial low-energy excitations~\cite{PhysRevLett.102.017205}. In this magnet, strong spin-orbit coupling (SOC) generates bond-dependent anisotropic exchange interactions. Combined with crystal electric field effects and strong electronic correlations, SOC induces $J_\text{eff}=1/2$ moments in an octahedral environment. An archetypal Kitaev magnet is Na$_2$IrO$_3$, in which the nearest IrO$_6$ octahedra share one of their edges and the Ir$-$O$-$Ir bond angle of around 90$^{\degree}$~\cite{hwan2015direct}. In such an octahedral environment, the electron hopping along the Ir$-$O$-$Ir pathways becomes orbital nonconserving, and as a result, the exchange coupling appears with discrete symmetry. Moreover, destructive quantum interference between the two Ir$-$O$-$Ir hopping paths across each edge-shared interface suppresses the Heisenberg interaction~\cite{PhysRevLett.102.017205}. In contrast, an Ising-like ferromagnetic interaction develops between $J_\text{eff}=1/2$ Ir$^{4+}$ moments, with an easy-axis anisotropy perpendicular to the Ir$-$O$-$Ir plane. This bond-dependent anisotropy imposes an extensive degeneracy due to the local constraints on each spin site, giving rise to extensive ground-state degeneracy and, consequently, exchange frustration in this bipartite lattice. The engineering of Kitaev spin liquids in honeycomb lattices is of paramount importance because they are potential contenders to host the most anticipated Majorana fermions, which manifest from the fractionalization of spin$-1/2$ moments~\cite{kitaev2006anyons}. While strong spin–orbit coupling plays a crucial role in stabilizing Kitaev-type magnetic interactions, it can also lead to other exotic quantum phases, including the realization of unconventional superconductivity with nontrivial pairing mechanisms~\cite{nelson2004odd}.

Beyond 3$d$/5$d$ transition metal, Kitaev model can also be realized on $d^7$ ions with an electronic configuration $t_{2g}^5e_g^2$ ($S=3/2,~L=1$)~\cite{PhysRevB.97.014407}, specifically for the Co$^{2+}$ or Ni$^{3+}$ ions in an octahedral environment as shown in Fig.~\ref{CrystalStructure}(c, d). Unlike many $3d$ ions, where the orbital moments are quenched due to a strong crystal electric field, the partially filled $t_\text{2g}$ orbitals in $\text{Co}^{2+}$ can retain an effective angular momentum $L_\text{eff}=1$. As a result, the spin-orbital multiplets comprise $12$ degenerate states. With the finite spin-orbit interaction ($\lambda \textbf{L}.\textbf{S}$), the twelve degenerate states further split into $J_\text{eff}=1/2, 3/2, $ and $ 5/2$ manifolds, as depicted in Fig.~\ref{CrystalStructure}(d)~\cite{PhysRevB.97.014407,kim2022spin}. In the high-spin Co$^{2+}$ state, the $e_g$ orbitals are half-filled, and there is one hole in the $t_{2g}$ state, which provides the spin degree of freedom for the realization of $J_\text{eff}=1/2$, with $e_g$ electrons forming a singlet~\cite{PhysRevB.97.014408}. Unlike the Kitaev magnet based on $4d^5/5d^5$, where the exchange path mainly involves $t_{2g}-t_{2g}$ electron hopping, the $J_\text{eff}=1/2$  Co$^{2+}$ moments interact via various exchange paths, such as $t_{2g}-t_{2g}$, $t_{2g}-e_{g}$ and $e_{g}-e_{g}$ hopping process mediated by the $2p$-orbital of the oxygen ligand~\cite{kim2022spin}. The ferromagnetic Kitaev exchange is derived predominately from the $t_{2g}-t_{2g}$ channel, whereas the $e_{g}-e_{g}$ and $t_{2g}-e_{g}$ hopping channels contribute to the Heisenberg and off-diagonal anisotropic exchange interactions, which tend to partially cancel each other~\cite{PhysRevB.97.014407,PhysRevLett.125.047201,PhysRevB.97.014408}. The ratio ($U/\Delta_{pd}$) between the on-site Coulombic interaction $U$ and the charge-transfer gap $\Delta_{pd}$, is comparatively higher in cobaltates than in $4d/5d$-based Mott insulators~\cite{PhysRevLett.125.047201}. The non-Kitaev exchanges, such as the Heisenberg coupling, are strongly affected by the value of $U/\Delta_{pd}$.

In addition to the spin-orbit-coupled bond-dependent anisotropic exchange Kitaev honeycomb lattice, the pure Heisenberg honeycomb lattice can also host exotic quantum phenomena because of its low connectivity and dimensional constraints. The Heisenberg model on the honeycomb lattice hosting antiferromagnetic exchange interactions $J_2=J_3=J_1/2$, where $J_1, J_2,$ and $J_3$ are the first, second, and third nearest-neighbor exchange couplings, respectively,  which is predicted to be highly frustrated with a macroscopically degenerate ground state~\cite{PhysRevLett.117.167201}. Although the honeycomb lattice comprises two interpenetrating triangular sublattices, a buckled honeycomb arises when these two sublattices are displaced relative to each other along the crystallographic $c$-axis. Some prototypical examples of buckled honeycomb magnets include FeP$_3$SiO$_{11}$~\cite{PhysRevB.110.184402}, KNiAsO$_4$~\cite{PhysRevResearch.5.013022}, Fe$_4$Nb$_2$O$_9$~\cite{PhysRevMaterials.4.084403}, etc. Honeycomb magnets are highlighted as potential candidates to host intriguing phenomena, including the magnetoelectric effect, colossal magnetoresistance, spintronics, and the magnetocaloric effect. Recently, the transition metal-based series A$_4$B$_2$O$_9$ (A = Co, Mn, Fe; B = Nb, Ta) has garnered significant attention due to its magnetoelectric properties~\cite{fischer1972new,PhysRevB.97.161106,PhysRevB.98.134438}. Except for the Ni$_4$Nb$_2$O$_9$ compound, all other compounds crystallized in the same trigonal space group $P\bar{3}c1$, and host magnetoelectric coupling~\cite{tailleur2020lack}. In Co$_4$Nb$_2$O$_9$ (CNO), Co$^{2+}$ ions form layers of buckled honeycomb structure along the crystallographic $c$-axis~\cite{fang2014large,PhysRevB.102.174443}. The neutron diffraction experiment on the single crystals of CNO reveals the presence of long-range magnetic ordering at $T_\text{N}=27 $ K with easy-plane magnetic anisotropy~\cite{PhysRevB.102.174443}. 
A large magnetoelectric coupling near the AFM ordering was observed due to the spin-flop-like transition—i.e., a field-induced reorientation of spins in a collinear antiferromagnet when an external magnetic field is applied parallel to the easy-axis, thereby minimizing the competing exchange, anisotropy, and Zeeman energies~\cite{PhysRevB.75.094425}—occurring at a critical magnetic field of magnitude $\sim 1.2$ T for the polycrystalline samples~\cite{kolodiazhnyi2011spin} and a much lower field of 0.2 T along the $[1\bar{1}0]$ direction in the single crystal samples~\cite{PhysRevB.93.075117}. The observed linear ME effect in CNO is possibly due to critical spin fluctuations near the Néel temperature~\cite{yin2016colossal}, while it has also been attributed to magnetic-order–driven symmetry lowering of the magnetic structure~\cite{PhysRevB.93.075117}. Substitution of a non-magnetic ion at the magnetic site provides an effective route to tune competing exchange interactions, modify spin–lattice coupling, and engineer ME responses in frustrated magnets~\cite{w6j5-ljfj,PhysRevX.5.031034}. With Mg substitution in Co$_4$Nb$_2$O$_9$ (yielding Co$_3$MgNb$_2$O$_9$), the Néel temperature 
$T_{\rm N}$ is suppressed from 27 K to 19 K, while the ME coupling is enhanced by nearly a factor of two compared to the parent compound~\cite{li2018enhancing}. Similarly, partial substitution of non-magnetic Mg ions in Fe$_2$Mo$_3$O$_8$ strengthens the spin-lattice coupling, thereby enhancing the ME coupling and stabilizing the two metamagnetic states. In this context, the A$_2$B$_2$O$_9$ and related series of buckled honeycomb magnets provide a rich platform for the experimental realization of intriguing magnetic states arising from spin–orbit coupling (SOC), competing exchange interactions driven by structural distortions, non-trivial low-energy excitations, and magnetoelectric coupling. Understanding the interplay between competing degrees of freedom, structural distortions, correlated disorder, and the resulting exchange hierarchy is crucial in honeycomb magnets for the rational design of quantum materials with the potential to host exotic quantum and topological many-body phenomena.
  
Herein, we report the crystal structure, magnetization, magnetocaloric effect (MCE), and specific heat studies of a distorted honeycomb lattice material, Co$_3$ZnNb$_2$O$_9$ (CZNO). Magnetization data show an anomaly at 14 K suggesting a phase transition. A $\lambda-$type anomaly in the temperature dependence of specific heat at $T=$ \~14 K, corroborating the anomaly in magnetic susceptibility and suggesting the onset of long-range antiferromagnetic ordering. 
The unusual behavior of the magnetic specific heat, $C_m\sim T^{1.45}$, well below the magnetic ordering temperature, indicates the presence of exotic low-energy excitations. Furthermore, the reduced entropy release in the ordered state, compared to the expected value, suggests a highly degenerate ground-state manifold. The special geometry of the buckled honeycomb structure of Co$^{2+}$ moments with edge-sharing octahedral environments and reduced local crystal field symmetry enhance the spin-orbit driven anisotropy that possibly induces competing exchange network in this distorted spin-lattice, and induces structural disotiortion. 
Also, the magnetization isotherm at low temperatures shows a spin-flop–like metamagnetic transition near 1.2 T. Interestingly, this field scale also corresponds to an anomaly in the dielectric response $\varepsilon'$, suggesting that the field-induced spin reorientation directly influences the electric degrees of freedom and provides evidence for magnetoelectric coupling. The pronounced low-temperature MCE response possibly associated with the modulation of distorted exchange network under external magnetic field further positions this material as a promising energy-efficient candidate for advanced cryogenic refrigeration applications.

 \begin{figure*}[ht]
\includegraphics[width=0.99\textwidth]{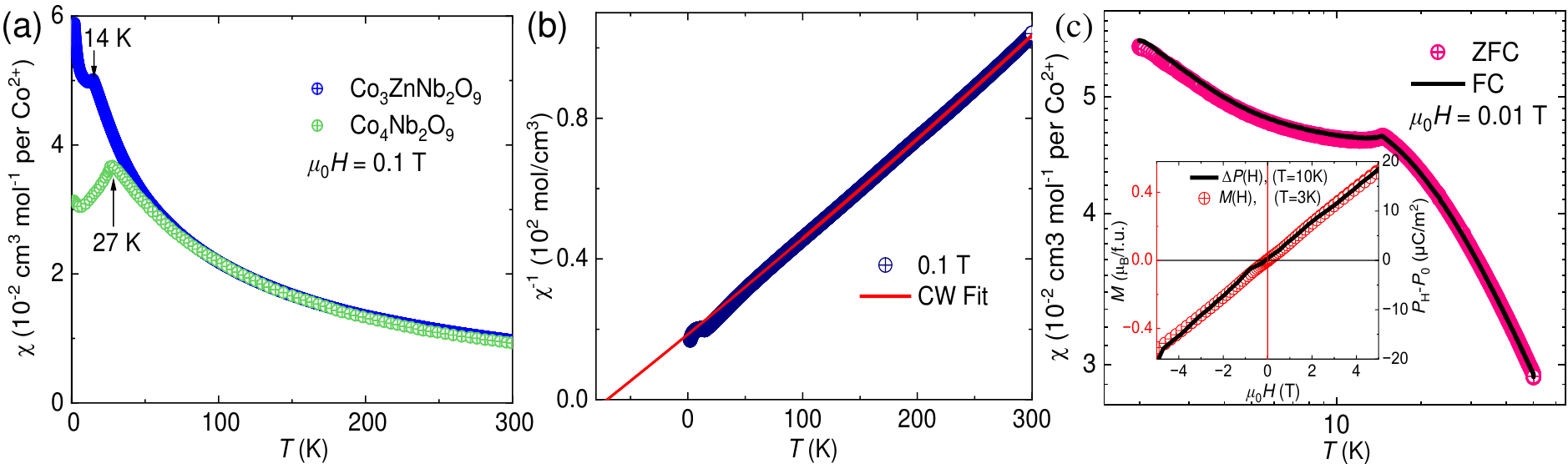}
  \caption{(a) A comparison between the temperature dependence of magnetic susceptibility $\chi(T)$, of Co$_3$ZnNb$_2$O$_9$ and Co$_4$Nb$_2$O$_9$, measured under an applied magnetic field of 0.1 T. The $\chi(T)$ of CNO reproduced with permission from ref.~\cite{PhysRevB.102.174443}. (b). The inverse magnetic susceptibility $\chi^{-1}(T)$ for CZNO is fitted to the Curie-Weiss law. (c) ZFC and FC magnetic susceptibilities measured under an applied field of 100 Oe; the inset shows linear behavior of polarization and magnetization at temperatures well below the antiferromagnetic transition. The polarization data is reproduced with
permission from ref.~\cite{martin2024compositional}.}
\label{Magnetization}
\end{figure*}

\section{Experimental details} 
The polycrystalline sample of Co$_3$ZnNb$_2$O$_9$ (CZNO) was prepared using the standard solid-state reaction method. Stoichiometric quantities of Co$_3$O$_4$ (Puratronic, 99.9985$\%$), ZnO (Alfa Aesar, 99.9995$\%$), and Nb$_2$O$_5$ (Alfa Aesar, 99.95$\%$) were mixed and thoroughly ground in a mortar and pestle. The homogeneous mixture was pelletized and sintered at temperatures between 900°C and 1000°C for 24 hours to 48 hours, with two intermediate grindings to ensure better homogeneity. The phase purity of the polycrystalline CZNO sample was examined using an Aeris PANalytical X-ray diffractometer with Cu-K$_\alpha$ radiation ($\lambda=1.5406~$\AA) at room temperature. Rietveld refinement of X-ray diffraction (XRD) data was performed using FullProf software. The magnetic measurement was conducted using a vibrating sample magnetometer (VSM) option attached to a physical properties measurement system (QD PPMS, USA) over a temperature range of 2 K $\leq$ $T$ $\leq$300 K, and specific heat measurement was carried out using PPMS over a temperature range of 2 K $\leq$ T $\leq$ 200 K and in various magnetic fields up to 9 T.
\section{Results and discussion}
\subsection{Crystal structure}

Rietveld refinement analysis of powder XRD diffraction data suggests that Co$_3$ZnNb$_2$O$_9$ crystallizes in the trigonal space group $P3\bar{c}1$. Figure~\ref{CrystalStructure}(a) shows the Rietveld refinement of XRD pattern recorded at room temperature, where a small fraction ($\sim 1.9~\% $) of non-magnetic impurity phase, ZnNb$_2$O$_6$, is found. Such a non-magnetic impurity is often unavoidable in polycrystalline samples \cite{PhysRevB.97.161106,PhysRevB.83.174412} and similar non-magnetic impurity peaks have also been reported in other Zn-doped compounds without any structural transition~\cite{PhysRevMaterials.7.014407}. The refined crystallographic parameters are provided in Table.~\ref{Table_1} that closely agree with those reported for single-crystals of isostructural Mn$_3$ZnNb$_2$O$_9$~\cite{rohweder1988kristallchemie}. Upon substituting Co$^{2+}$ at Mn$^{2+}$ site, the lattice parameters are observed to decrease slightly. A similar trend has also observed in the single crystal studies, where the hexagonal lattice contracts modestly, consistent with the systematic decrease of the lattice parameters $a$ and $c$ observed when going from Mn$_4$Nb$_2$O$_9$ to Co$_4$Nb$_2$O$_9$~\cite{PhysRevB.102.174443,rohweder1988synthese}. Each Co$^{2+}$ ion is coordinated by six oxygen ions in a CoO$_6$ unit within the $ab$-plane.
The magnetic ion Co$^{2+}$ occupies a trigonally distorted octahedral site, and the Co$^{2+}$ moments form a buckled honeycomb network of Co1-Co2 in the $ab-$plane (see Fig.~\ref{CrystalStructure}(b) and (c)). Strong electrostatic repulsion between these cations drives them apart, distorting the oxygen framework and causing the Co honeycomb layers to ripple or ``buckle'' out of the plane rather than remaining flat. 
The first nearest neighbor bond length, Co1-Co2, equals 2.99(7)$\text{\AA}$ within the $ab$ plane, which constitutes the honeycomb lattice, while along the $c$ axis, the Co1-Co3 ions form dimers with a bond length of 3.008(2) $\text{\AA}$.The bond angle Co1-O-Co2 in the honeycomb plane is close to 90.1(3)$^{\circ}$. Such nearly $90^{\circ}$ edge-sharing octahedral geometries are a key structural ingredient for realizing bond-directional Kitaev-like exchange in systems with strong SOC, as observed in several $4d/5d$ honeycomb magnets~\cite{PhysRevLett.102.017205}. The trigonal distortion of CoO$_6$ octahedra arises from the displacement of Co$^{2+}$ away from the octahedral center, which is analogous to the behavior reported for Mn$^{2+}$ sites in the Mn$_3$ZnNb$_2$O$_9$~\cite{rohweder1988kristallchemie}. In this honeycomb material, the trigonally compressed CoO$_6$ octahedra lift the orbital degeneracy and can generate an effective spin-orbit-entangled Kramers doublet, so that the local distortion and connectivity of CoO$_6$ units are expected to play a decisive role in stabilizing anisotropic exchange interactions thereby determining the magnetic ground state. Structural analysis of CZNO further reveals octahedral distortions associated with the face- and edge-sharing polyhedra along $c-$axis and $ab-$plane, respectively (see Fig.~\ref{CrystalStructure}(b)). Such distortions lower the local symmetry from a cubic environment, modify the exchange pathways and possibly induce anisotropy in exchange interactions that play a crucial role in stabilizing non-trivial ground state of this bipartite lattice~\cite{PhysRevLett.102.017205,PhysRevB.97.014407}.

\begin{table}[htb]
\caption{Rietveld refined unit cell parameters: $a$ = $b$ = 5.1808(2) \AA, $c$ = 14.1378(1) \AA, $\alpha=\beta=90\degree$,$\gamma=120\degree$. The goodness of fit parameter chi-square ($\chi^2$), weighted profile residual (R$_{wp}$), profile residual (R$_p$), and experimental profile (R$_{exp}$) values are 2.39, 1.21$\%$, $0.78\%$, and $0.77\% $, respectively. The atomic parameters of the CZNO are given below.}
\label{Table_1}
\begin{ruledtabular}
\begin{tabular}{ccccccc}
Atom & Site & $x$ & $y$ & $z$ & $U_{iso}$ & Occ. \\[0.1 in]
\hline
Nb1 &$2a$ & 0.000 & 0.000 & 0.3589(5)&0.032 & 1 \\[0.1 in]
Nb2 &$2a$ & 0.000 & 0.000 & 0.137(1)&0.047 & 1 \\[0.1 in]
Zn1& $2b$&0.333(3)&0.666(6)& 0.304(4)&0.006 &1\\[0.1 in]
Co1& 4$2b$&0.333(3)&0.666(6)& 0.0123(1)&0.002&1\\[0.1 in]
Co2&4$2c$&0.666(6)&0.333(3)&0.485(8)&0.002
&1\\[0.1 in]
Co3&4$2c$&0.666(6)&0.333(3)&0.199(4)&0.002
&1\\[0.1 in]
O1 & 6$d$&0.353(1) &0.321&0.087&0.002
&1\\[0.1 in]
O2&6d&0.32&0.336&0.414&0.002
&1\\[0.1 in]
O3&6d&0.264&0.966&0.243&0.002
&1\\
\end{tabular}
\end{ruledtabular}
\end{table}


\subsection{Magnetic susceptibility}

 To investigate the magnetic properties, the temperature-dependent dc magnetic susceptibility ($\chi(T)$) was recorded in the temperature range of $2-300$ K with an applied magnetic field of strength 0.1 T. As shown in Fig.~\ref{Magnetization}(a), CZNO shows a cusp around 14 K, which is attributed to the onset of long-range magnetic ordering. The observed long-range ordering is consistent with previous results~\cite{martin2024compositional}. In comparison, the parent compound CNO exhibits long-range ordering around 27 K~\cite{PhysRevB.94.094427,PhysRevB.102.174443}. This suggests that 25\% doping of Zn at the Co-site has suppressed the ordering temperature by a factor of two compared to the original material. Upon further cooling below $T_\text{N}$, an upturn in $\chi(T)$ which is not common in many antiferromagnets. The upturn arises from uncompensated spin moments or defects due to site dilution that disrupts the compensation between magnetic sublattices. We performed Curie–Weiss (CW) fitting of the inverse magnetic susceptibility data in the temperature range $100-300$ K (Fig.~\ref{Magnetization}(b)) using the relation $
\chi(T) = \chi_0 + \frac{C}{(T-\theta_{\rm CW})},
$ where $\chi_0,$ $C$, and $\theta_{\rm CW}$ are the temperature-independent susceptibility, Curie constant, and Curie–Weiss temperature, respectively. The fitting yields a characteristic $\theta_\text{CW}\sim-70$ K, $C=3.847(3)~{\rm cm^3\,K\,mol^{-1}}$, and $\chi_0=7.31(7)\times10^{-4}~{\rm cm^3\,mol^{-1}}$. The temperature-independent susceptibility consists of core diamagnetic ($\chi_\text{dia}$) and Van-Vleck paramagnetism ($\chi_\text{VV}$). For CZNO, the calculated $\chi_\text{dia}$ is $-1.77\times 10^{-4}$ cm$^3$/mol and the $\chi_\text{VV}=\chi_0-\chi_\text{dia}\approx 9.08\times10^{-4}$ cm$^3$/mol. The large and negative Curie-Weiss temperature $\theta_\text{CW}=-70$ K indicates the presence of dominant antiferromagnetic interactions among the Co$^{2+}$ moments. The effective magnetic moment was extracted from the Curie constant according to $\mu_\text{eff}=\sqrt{8C}\mu_\text{B}$, resulting in $\mu_\text{eff}\approx5.54\mu_\text{B}$. This value of effective moment is comparatively higher than the expected $\mu_{\mathrm{eff}}$ of high-spin Co$^{2+}$ ($d^7, S = 3/2$), which is $3.87~\mu_\text{B}$. The effective moment value deviates significantly from the spin-only moment ($2.86~\mu_B$), indicating the roles of the crystal electric field and spin-orbit coupling. In an ideal octahedral field, the Co$^{2+}$ moment adopts a $t_\text{2g}^5e_g^2$ configuration for $S=3/2$ (see Fig.~\ref{CrystalStructure}(d)).
An enhanced $\mu_{\mathrm{eff}}$ suggests the presence of unquenched orbital angular momentum in the CZNO compound, consistent with other Co$^{2+}$-based antiferromagnets~\cite{zvereva2016orbitally}. The trigonal distortion of CoO$_6$ away from the octahedral center lowers the local symmetry from $O_h$ and splits the $t_{2g}$ level into an $a_{1g}$ singlet and $e_{g}^{'}$ doublet. As depicted in Fig.~\ref{Magnetization}(c), zero-field-cooled (ZFC) and field-cooled (FC) magnetization
data measured under an applied magnetic field of 0.01 T show no significant divergence, which rules out spin glass transition below $T_\text{N}$. A five-quadrant M-H curve recorded at 3 K (inset of Fig.~\ref{Magnetization}(c)) shows the absence of any hysteresis in the limit of instrument resolution. The absence of hysteresis in the isothermal magnetization at 3 K ( See Fig. \ref{Magnetization}(c)) rules out the presence of a ferromagnetic component. Furthermore, the linear behavior of the magnetization at 3 K up to 9 T suggests dominant antiferromagnetic interactions between the Co$^{2+}$ moments.

In CZNO, the edge-sharing CoO$_6$ geometry with Co--O--Co $\approx 90.1(8)^{\circ}$ has profound implications for determining the magnetic ground state. Theoretical studies have demonstrated that in edge-sharing CoO$_6$ octahedra, spin-orbit coupling can generate bond-directional exchange interaction between effective pseudospin-1/2 moments. The buckled honeycomb structure in CZNO further relaxes the symmetry constraints, which may allow multiple competing anisotropic interaction paths~\cite{PhysRevB.94.094427}. Despite the 90$^{\circ}$ bond angle, the net exchange is antiferromagnetic in nature as $\theta_\text{CW}\approx-70$ K. In Co$_3$ZnNb$_2$O$_9$, deviations of the CoO$_6$ octahedra from the underlying ideal symmetry, arising from correlated but unavoidable structural distortions~\cite{keen2015crystallography}, modify the Co–O–Co superexchange routes and induces an anisotropic hierarchy of exchange interactions. This partially release the geometric frustration inherent to the Co$^{2+}$ honeycomb network and consequently stabilizes antiferromagnetic ordering at 14 K. The frustration parameter $f=|\theta_\text{CW}|/T_\text{N}= 5$ suggests that CZNO is a moderately frustrated spin-orbit coupled antiferromagnet. 
 \begin{figure*}[ht]
\includegraphics[width=0.75\textwidth]{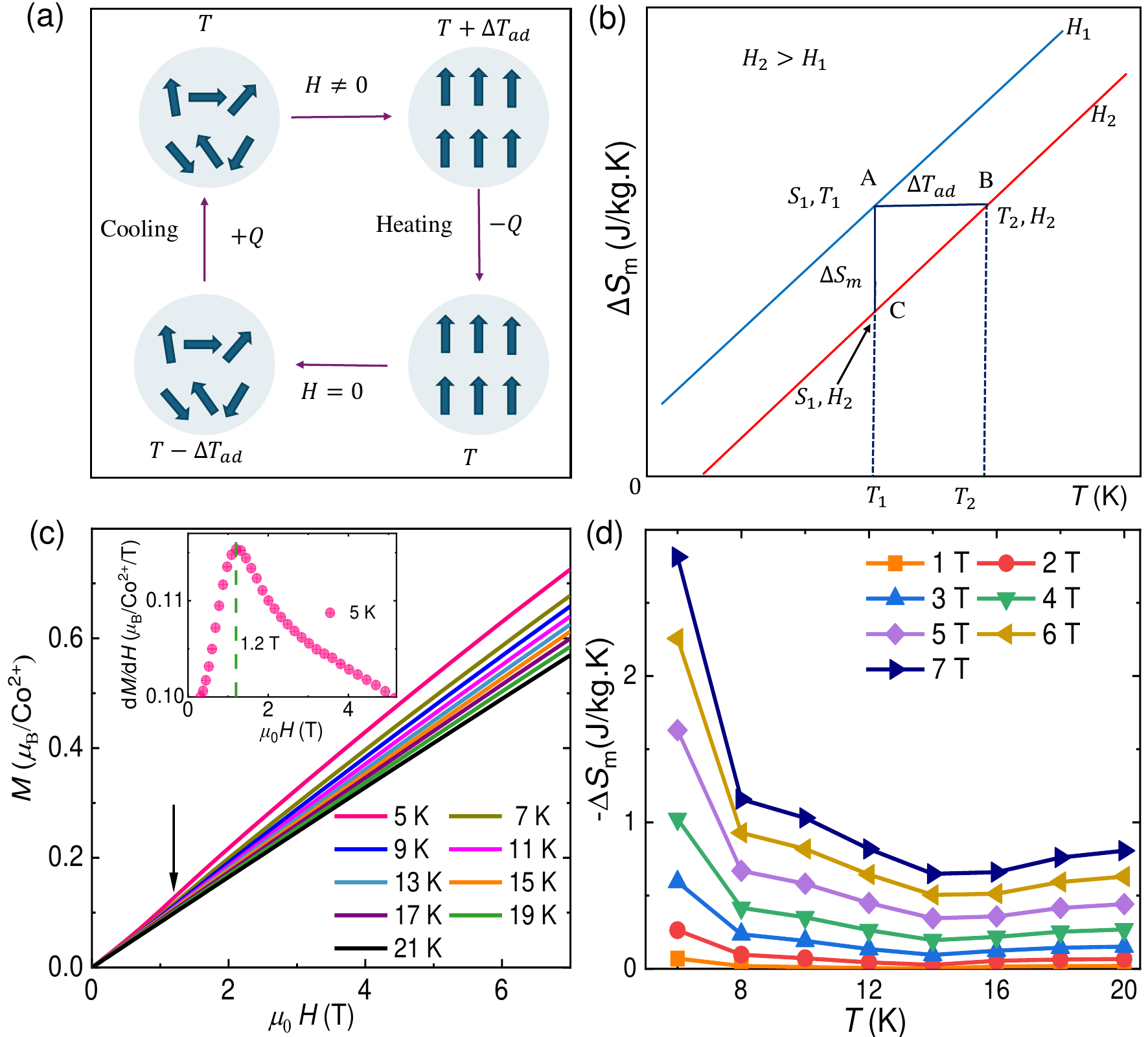}
  \caption{(a) The schematic diagram of the magnetocaloric refrigeration cycle, where two steps are involved: an adiabatic step in which the entropy remains constant and temperature changes and an isothermal process that induces a change in entropy while maintaining a constant temperature~\cite{nair2018magnetocaloric}. (b) The schematic diagram of entropy as a function of temperature and magnetic field, the vertical line represents the isothermal entropy change ($\Delta S_m$), while the horizontal line denotes the adiabatic temperature change ($\Delta T_\mathrm{ad}$)~\cite{pecharsky1999magnetocaloric}. (c) Isothermal magnetization at various temperatures in the first quadrant up to a magnetic field of 7 T. The inset shows derivative of magnetization with respect to magnetic field at 5 K as a function of the magnetic field. (d) The entropy change as a function of temperature for various magnetic fields derived from magnetization isotherms recorded at different temperatures.}
\label{Magnetocaloric}
\end{figure*}
Frustration in this bipartite lattice may be attributed to the presence of competing exchange interactions within the host honeycomb spin lattice.

The basic principle of magnetic refrigeration is illustrated in Fig.~\ref{Magnetocaloric}(a), where the application and removal of an external magnetic field ($H$) drive an adiabatic process. This cycle harnesses the entropy difference between aligned and disordered spin states, causing the material to release heat ($Q$) upon magnetization and absorb lattice heat during demagnetization.
Considering the field-dependent magnetization, the spin-flop transition, and the absence of measurable magnetic hysteresis in the isothermal magnetization measurements, we expect the presence of MCE in polycrystalline CZNO. The MCE refers to the change in temperature ($\Delta T_{ad}$) of a material when subjected to an external magnetic field and is quantified by the isothermal magnetic entropy change as the magnetic field is varied (see Fig.~\ref{Magnetocaloric}(b)). 
In antiferromagnetic systems, the application of a magnetic field can modify the balance between exchange interactions and magnetic anisotropy. This may lead to a progressive reconfiguration of the spin structure and a concomitant reduction of magnetic entropy. The field-induced suppression of spin correlations is particularly effective near the magnetic phase transition, where spin fluctuations are more pronounced.
In order to quantify MCE in the titled honeycomb material, we measured the isothermal magnetization for MCE analysis at 2 K intervals from 5 K to 21 K in magnetic fields up to 7 T (see Fig.~\ref{Magnetocaloric}(c)). A clear deviation from linear M–H behavior is observed for 5 K, which is well below the $T_{\rm N}$, with the magnetization exhibits noticeable nonlinearity above ~1.2 T. As shown in the inset of Fig.~\ref{Magnetocaloric}(c), the $dM/dH$ shows a pronounced peak around 1.2 T in the magnetization data measured at 5 K, which is below $T_\text{N}$. The peak at 1.2 T correspond to the critical field ($H_\text{sf}$), indicative of a spin-flop–like metamagnetic transition~\cite{kolodiazhnyi2011spin}. The isothermal magnetic entropy change, $\Delta S_m$, near the transition temperature was calculated from the M-H data using Maxwell's thermodynamic relations based on the isothermal magnetization experiments, as given by the following relation~\cite{wu2022magnetic}.

\begin{equation}
\begin{aligned}
-\Delta S_m (T,H) & = \int_{0}^{H} \left(\frac{\partial S}{\partial H}\right)_T \, dH = \int_{0}^{H} \left(\frac{\partial M}{\partial T}\right)_H \, dH \\
&= \sum_0^H \left( \frac{M(T_1)-  M(T_2)}{T_1-T_2} \right) \, \Delta H
\end{aligned}
\end{equation}

Where $M(T_1)$ and $M(T_2)$ are the magnetizations recorded at temperatures $T_1$ and $T_2$, respectively, for a magnetic field change $\Delta H$. Fig.~\ref{Magnetocaloric}(a) and~\ref{Magnetocaloric}(b) depict schematic representations of the MCE. To determine the isothermal entropy change, we employed a two-step calculation process. First, we computed the difference in magnetic moments between two adjacent temperatures. This difference was then divided by the temperature interval between the two consecutive temperatures.
Finally, we integrated the resulting values with respect to magnetic fields. Fig.~\ref{Magnetocaloric}(d) shows the temperature dependence of isothermal entropy changes across $T_{\rm N}$. The maximum change in isothermal magnetic entropy is 2.81 J/kg · K for a field variation from 0 to 7 T at 14 K. The observed MCE in CZNO is relatively small due to strong antiferromagnetic interactions in CZNO~\cite{zhang2024crystal}. In antiferromagnetic materials, the alignment of magnetic moments on neighboring atoms in opposite directions leads to the cancellation of net magnetization, resulting in a smaller change in magnetization under an applied magnetic field compared to ferromagnetic materials, thereby reducing the MCE. The isothermal magnetization measurements revealed that a high magnetic field is required to saturate the magnetic moment of Co$^{2+}$ ions in CZNO ~\cite{zhang2024crystal}. The small change in isothermal magnetic entropy is also found in other antiferromagnetic systems~\cite{sahoo2018influence,meng2022negative} and  Table \ref{Table_3} shows the isothermal entropy change of a few representative frustrated magnets.

\begin{table*}[htb]
\caption{The isothermal entropy change of a few frustrated magnets.}
\label{Table_3}
\begin{ruledtabular}
\begin{tabular}{ccccccc}
Compounds &Space group  & $T_{\rm N}/T_{\rm C}$ (K) & $\theta_{\rm CW}(\mathrm{K})$& $\mu_0 H(\mathrm{T})$ & $-\Delta S_m \mathrm{max} (\mathrm{J/Kg.K})$ & [ref.] \\[0.1 in]
\hline


FeMnO$_3$& $Ia3$&36&-69& 9& 1.5& \cite{rayaprol2015magnetic}\\[0.1 in]

GdInO$_3$&$P63cm$&2.1
&-7.41&7&18.37&\cite{wu2022magnetic}\\[0.1 in]

 Ho$_2$CoMnO$_6$ & $P2_1/n$&76&79.4& 7& 13.5&\cite{PhysRevMaterials.6.104401}\\[0.1 in]
Co$_3$ZnNb$_2$O$_9$&$P\bar{3}c1$ &14 &-70&7&2.81&\textbf{This work}
\end{tabular}
\end{ruledtabular}
\end{table*}

\subsection{Specific heat and magnetoelectric behavior}

 \begin{figure*}[t]
\includegraphics[width=0.85\textwidth]{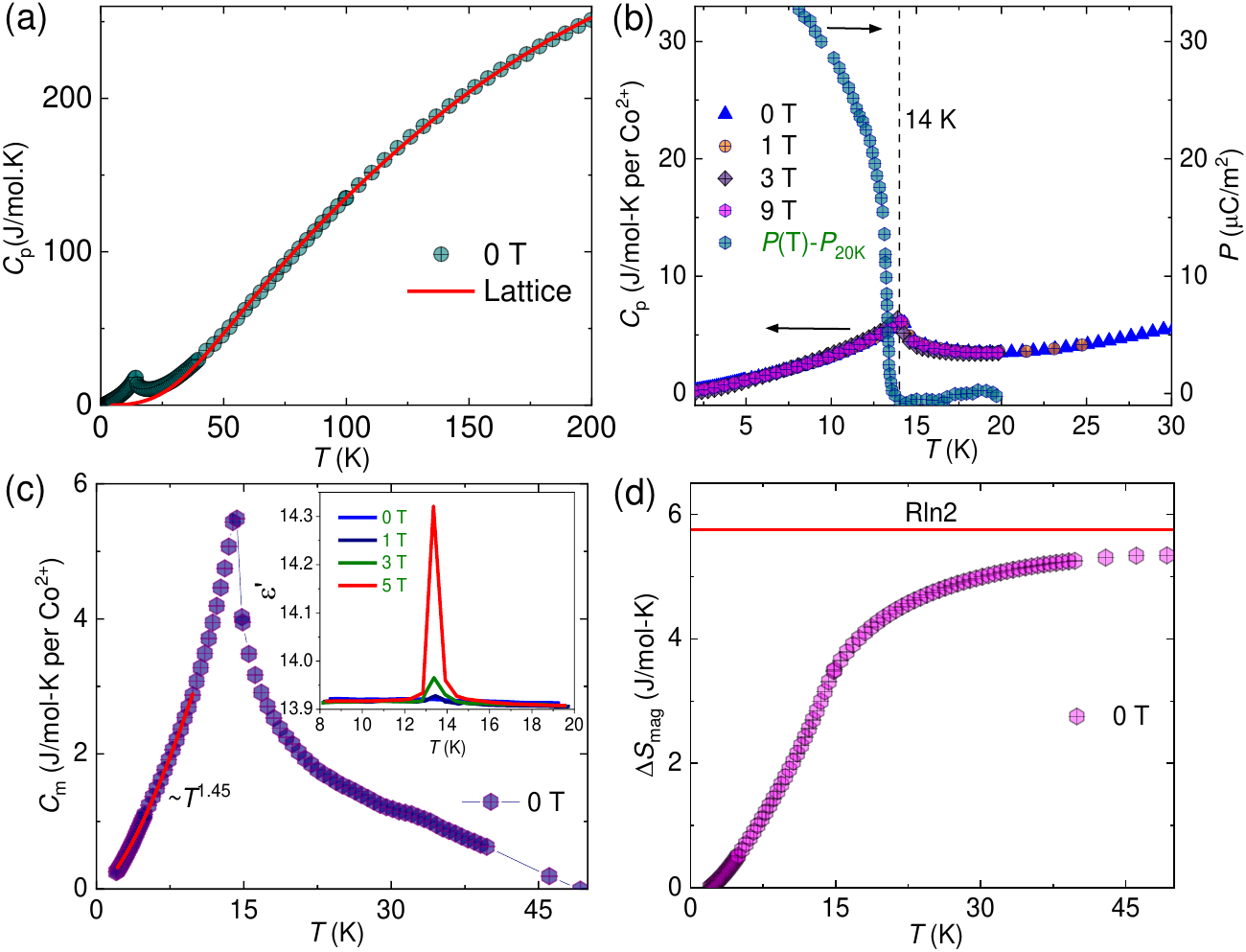}
  \caption{(a) Temperature dependence of specific heat in zero magnetic field. The solid red line represents the lattice contribution arising from phonon excitations. (b) The field dependence of specific heat was measured up to 9 T and showing almost no significant shift in $T_\text{N}$. This result is compared with the electric polarization data adapted from ref.~\cite{martin2024compositional}. (c) The magnetic specific heat of CZNO at zero field, obtained after subtracting the lattice contribution using Debye-Einstein fitting, and the red line shows the power law behavior of $C_\text{m}$ with temperature below $T_\text{N}$. A sharp $\lambda-$type anomaly at 14 K, indicates the onset of long-range magnetic ordering. In the inset, we show the dielectric permittivity $\varepsilon'$ measured under various applied fields, reproduced with permission from ref.~\cite{martin2024compositional}. (d) Magnetic entropy obtained from the corresponding $C_\text{m}$ for zero field specific heat result.}
\label{specific heat}
\end{figure*}

Specific heat is a sensitive probe for tracking exotic states and the associated low-energy excitations in frustrated honeycomb magnets. To probe the nature of the ground state, specific heat ($C_p$) measurements were performed down to 2 K in magnetic fields of up to 9 T. Fig.~\ref{specific heat}(a) represents the temperature-dependent specific heat $C_p(T)$ in the range 2 K $\le T \le 200$ K in zero field. A sharp $\lambda$-type anomaly is observed at $\sim 14$ K associated with the onset of long-range antiferromagnetic ordering, consistent with the temperature-dependent magnetic susceptibility data. The field-dependent specific heat data, as shown in Fig.~\ref{specific heat}(b), measured up to an applied field of 9 T, show no noticeable variation, which indicates that the long-range ordering is robust against the applied magnetic field. To extract the magnetic specific heat $C_\text{m}$ in the absence of a non-magnetic analog, we analyze the experimental data using the Debye-Einstein model to subtract the lattice components. In this model, the Debye term represents the contribution from low-energy acoustic phonon modes associated with collective lattice vibrations, while the Einstein terms capture higher-energy optical modes arising from localized atomic vibrations in the crystal lattice. Accordingly, the specific heat was fitted using a linear combination of one Debye and three Einstein terms~\cite{PhysRevB.90.035141}:
\begin{equation}
\label{DE equation}
\begin{aligned}
    C_{latt}(T)=C_D \left[ 9k_B \left(\frac{T}{\theta_D}\right)^3 \int_0^{\theta_D /T} \frac{x^4 e^x}{(e^x-1)^2} dx\right]  \\
    + \sum_{i=1}^3 C_{E_i} \left[ 3R \left( \frac{\theta_{E_i}}{T}\right)^2 \frac{e^{\theta_{E_i}/T}}{(e^{\theta_{E_i}/T}-1)^2} \right]
\end{aligned}
\end{equation}
Here $x=\frac{\hbar\omega}{k_B T}$, $R$ is the universal gas constant, $k_B$ is the Boltzmann constant, $\theta_D$ and $\theta_{E_i}$ are the characteristic Debye and Einstein temperatures, respectively. From the fitting we obtain the characteristic parameters as: $C_{D_1}=1$, $\theta_{D_1}=116$ K and three Einstein terms:  $C_{E_1}=4.5$, $\theta_{E_1}=225$ K, $C_{E_2}=3.5$, $\theta_{E_2}=779$, and $C_{E_3}=6$, $\theta_{E_3}=446$ K. This is consistent
with the total number of atoms per formula unit (15 in CZNO). The low-temperature magnetic specific heat follows a power-law behavior, $C_\text{m}\propto T^{1.45}$ (Fig.~\ref{specific heat}(c)). This marks a clear deviation from the conventional 3D antiferromagnetic ordering in CZNO, where $C_\text{m}\propto T^{3}$. Such nontrivial power-law behavior has been widely associated with frustration-induced soft modes, exchange randomness, or glassy spin dynamics, where a continuum of low-energy states emerges. 

Figure~\ref{specific heat}(d) displays the temperature-dependent magnetic entropy ($S_\text{mag}$) in zero field, which was calculated by integrating the magnetic specific heat by temperature ($C_m/T$) over the low-temperature range. It is observed that the magnetic entropy increases with increase in temperature and saturates near 45 K with a value of 5.3 J/mol·K, which is nearly equal to the expected value for the effective spin$-1/2$ system ($\approx5.76$ J/mol·K). 
This indicates that the low-energy degree of freedom in CZNO is possibly due to the $J_{\text{eff}}=1/2$ Kramers doublet arising from strong SOC and crystal-field effects on the Co$^{2+}$ ions. At $T_\text{N}$, only about 54\% of the total entropy $\text{Rln2}$ is released that suggest a highly degenerate manifold of low-energy excitations. For a conventional 3D antiferromagnet, the majority of the magnetic entropy is expected to be released in the vicinity of the transition temperature $T_\text{N}$. A substantial reduction in $S_m$ at $T_\text{N}$ suggests that a significant amount of spin entropy is quenched at high temperature, which could be due to the onset of short-range spin correlations above the transition temperature.

It is worth noting that certain antiferromagnets with broken time-reversal and inversion symmetries were proposed to host free energy terms of the form $\Phi=EH$~\cite{landau2013electrodynamics, dzyaloshinskii1960magneto}.
The dielectric response of CZNO provides a direct probe of the magnetoelectric coupling. As reported in the ref~\cite{martin2024compositional}, the zero-field data on dielectric permittivity $\varepsilon^{'}(T)$ remains featureless across the transition temperature $T_\text{N}$. This suggests that there is no spontaneous ferroelectric transition. However, with the application of the magnetic field, an anomaly appears in $\varepsilon^{'}(T)$ around $T_\text{N}$ (see inset of Fig.~\ref{specific heat}(c)). A $\lambda-$type anomaly appears in the measurement of $\varepsilon^{'}(T)$ at $T_\text{N}$, with increasing magnetic field, the amplitude of the anomaly increases significantly above an applied field of 1 T, as shown in the inset of Fig.~\ref{specific heat}(c).  Coincidentally, in CNO, the critical field $H_\text{sf}$ is also 1.2 T, and above this field, a narrow peak appears in the measurement of $\varepsilon'$ at the transition temperature $T_\text{N}$~\cite{kolodiazhnyi2011spin}. Notably, the anomaly in $\varepsilon'$, as evident in CZNO and CNO antiferromagnets, is confined to a very narrow temperature range across $T_\text{N}$. 
 
 This suggests that the onset of the dielectric anomaly differs from that of a structural or lattice-driven instability. This behavior in CZNO is characteristic of linear magnetoelectric coupling, which was first reported for Cr$_2$O$_3$~\cite{astrov1960magnetoelectric}. As shown in Fig.~\ref{specific heat}(b), the polarization develops below $T_\text{N}$ under an applied field of 9 T, reaching up to 33 $\mu$C m$^{-2}$ at 8 K, which further suggests that the electric response is magnetically driven. The absence of polarization below $T_\text{N}$, as evident from Fig.~\ref{specific heat}(b), rules out extrinsic effects such as Maxwell–Wagner polarization. In polycrystalline Eu$_2$CoMnO$_6$, a double perovskite, the dielectric anomaly observed at high temperature and at low frequencies is mainly attributed to Maxwell–Wagner type interfacial polarization, which is characterized by strong frequency dispersion, dielectric loss, and grain-boundary relaxation. In contrast, polarization in CZNO is intrinsically linked to magnetically driven spin–lattice coupling, which explains the absence of dielectric response in a zero magnetic field. Another key feature can be obtained from the field dependent magnetization $M(\text{H})$ and polarization $P(\text{H})$ at low temperatures. As depicted in the inset of Fig.~\ref{Magnetization}(b), the magnetization is linearly evolving with the field, which is characteristic of antiferromagnets, and the polarization is almost linear with the field. This further confirms that the magnetoelectric response is linear, and, the polarization can be expressed as $ P = \alpha_{i,j} H$, where $\alpha_{i,j}$ is the linear magnetoelectric tensor. Here, CZNO is a type-II multiferroic in which the polarization is typically low, and the spin-induced polarization accounts for the dielectric response under an applied magnetic field. In contrast, type-I multiferroics possess ferroelectricity above the ordered temperature and typically show larger electric polarizations. In general, the ordering temperature of type-I multiferroics lies at high temperature; however, the magnetoelectric coupling is usually stronger in the case of type-II multiferroics~\cite{mostovoy2024multiferroics}. The interplay between SOC driven anisotropy, enhanced by the reduced local crystal-field symmetry, and competing exchange interactions gives rise to a metamagnetic-like transition in the ordered state under a weak magnetic field. In addition, field-induced reconfiguration of the distorted exchange network in this frustrated magnet produces a magnetocaloric response in the ordered state.

\begin{figure}
    \centering
   \includegraphics[width=1\linewidth]{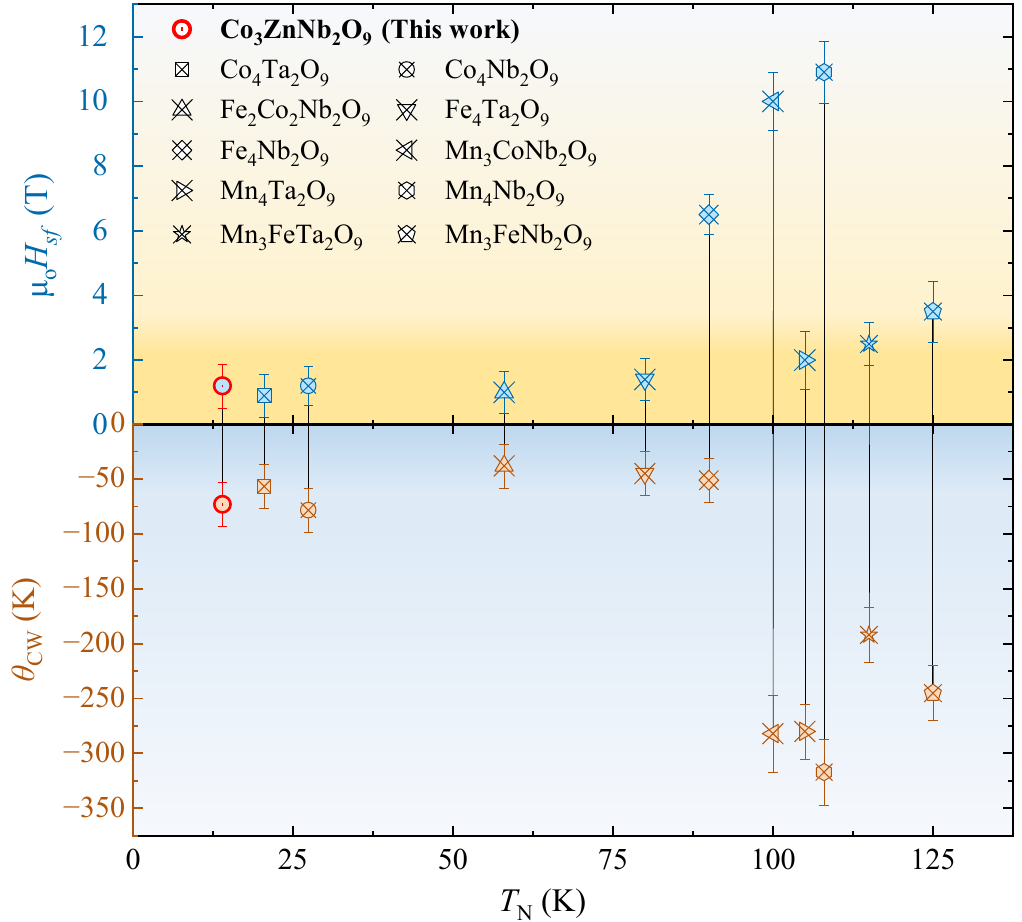}
    \caption{Correlated evolution of magnetic energy scales across the A$_4$B$_2\text{O}_9$ family ($A=\text{Mn, Fe, Co, Zn}$; $B=\text{Nb, Ta}$): \textbf{[Top]} Variation of the spin-flop transition field ($\mu_0 H_{sf}$) as a function of Néel temperature ($T_{\rm N}$), illustrating the relationship between magnetic anisotropy energy and thermal stability. \textbf{[Bottom]} The Curie-Weiss temperature ($\theta_{\rm CW}$) plotted against $T_{\rm N}$, highlighting the evolution of exchange coupling strength and magnetic frustration ($f = |\theta_{\rm CW}|/T_{\rm N}$) across the series~\cite{PhysRevB.94.094427,kolodiazhnyi2011spin,PhysRevB.97.161106,maignan2024enhancement,maignan2021fe}. The figure highlights a feasible scaling between anisotropy and exchange energy scales, consistent with $H_\text{sf}\propto\sqrt{H_EH_A}$, where the exchange field $H_\text{E}$ and $H_\text{A}$ set the ordering temperature $T_\text{N}$.} 
    \label{fig:PhaseDia}
\end{figure}

\begin{table*}[ht]
\centering
\caption{A comparative account of niobate series on a distorted honeycomb network with their magnetic ground state properties, highlighting the details of how non-magnetic substitution (Zn, Mg) systematically dilutes the magnetic sublattices, thereby enhancing in the magnetic frustration ($f = |\theta_{\rm CW}|/T_{\rm N}$) through the suppression of long-range ordering (LRO), and determining whether the ground state is antiferromagnetic (AFM) or ferrimagnetic (FIM).}
\label{tab:series}
\renewcommand{\arraystretch}{1.4} 
\begin{tabular}{l c c c c c c l}

\hline
\toprule
{Composition}~~~~~~~~~ & {Space Group}  & ~~~~~~~{$x$}~~~~~~ &~~~ {$T_{\rm N}$(K)}~~~&~~{$\theta_{\rm CW}$(K)}~~~& ~~ $f=|\theta_{\rm CW}|/T_{\rm N}$~~& ~~~~~~~{Ground State}~~~~~~~~& ~~~~{Ref.} ~~~~\\
\midrule

\multicolumn{5}{l}{{$[\text{Co}_{(1-x)}\text{Zn/Mg}_{(x)}]_4\text{Nb}_2\text{O}_9$ }} \\
\midrule
$\text{Co}_4\text{Nb}_2\text{O}_9$   &   $P\bar{3}c1$     & 0    & $27.4$ & $-78.3$ &  $\sim 3$ & AFM (LRO) & \cite{kolodiazhnyi2011spin,li2018enhancing} \\
\textbf{$\text{Co}_3\text{Zn}\text{Nb}_2\text{O}_9$} &   $P\bar{3}c1$  & 0.25 & $14$ & $-70$ &  $\sim 5$ &  AFM (LRO) & \scriptsize{\textbf{This Work}} \\

$\text{Co}_3\text{Mg}\text{Nb}_2\text{O}_9$ &    $P\bar{3}c1$   & 0.25 & $19$ &   - &  - & AFM (LRO)  & \cite{li2018enhancing} \\
$\text{Co}_2\text{Mg}_2\text{Nb}_2\text{O}_9$ &   $P\bar{3}c1$ &  0.50  & $11$ &  - &  - & AFM (LRO)  & \cite{li2018enhancing} \\
$\text{Co}\text{Mg}_3\text{Nb}_2\text{O}_9$  &   $P\bar{3}c1$ & 0.75  &   $<3$  & - &  - & No LRO down to 3 K  & \cite{li2018enhancing} \\

\midrule
\multicolumn{5}{l}{{$[\text{Mn}_{(1-x)}\text{Zn}_{(x)}]_4\text{Nb}_2\text{O}_9$}} \\
\midrule
$\text{Mn}_4\text{Nb}_2\text{O}_9$   &    $P\bar{3}c1$        & 0    & 125 & $-247$ &  $\sim 2$ &   AFM (LRO) & \cite{maignan2024enhancement,rohweder1988synthese} \\
$\text{Mn}_3\text{Zn}\text{Nb}_2\text{O}_9$  &   $P\bar{3}c1$ & 0.25 & 82  & $-179$ &  $\sim 2$ &  AFM (LRO) &  \cite{maignan2024enhancement,rohweder1988kristallchemie} \\
$\text{Mn}_2\text{Zn}_2\text{Nb}_2\text{O}_9$ &   $P\bar{3}c1$ & 0.50  & -  & $-104$ &  - &  -  & \cite{maignan2024enhancement,rohweder1988synthese} \\
\midrule
\multicolumn{5}{l}{{$[\text{Ni}_{(1-x)}\text{Mg}_{(x)}]_4\text{Nb}_2\text{O}_9$}} \\
\midrule
$\text{Ni}_4\text{Nb}_2\text{O}_9$    &    $Pbcn$        & 0    & $77\pm 1$ & $-215$  &  $\sim 3$ & FIM  & \cite{PhysRevB.52.9595,martin2024magnetic} \\
$\text{Ni}_3\text{Mg}\text{Nb}_2\text{O}_9$ &   $Pbcn$   & 0.25 & 45.5       & $-137.2$  &  $\sim 3$ & FIM & \cite{tarakina2007crystal} \\
$\text{Ni}\text{Mg}_3\text{Nb}_2\text{O}_9$   &    $P\bar{3}c1$  & 0.75 & $<2$ & $-13.1$ &  $> 6 $ &  No LRO down to 2 K & \cite{tarakina2007crystal} \\

\bottomrule
\hline
\end{tabular}
\end{table*}

\section{Discussion}



The titled material represents a rare example of a distorted honeycomb magnet in which the intriguing interplay of anisotropy—arising from spin–orbit–entangled Co$^{2+}$ moments in edge-sharing CoO$_6$-octahedra and exchange pathways that foster anisotropic, competing exchange interactions couples electric polarization, magnetic order, lattice symmetry, and magnetic entropy, thereby providing a promising route to realizing multiferroicity and a sizable magnetocaloric effect. In the paramagnetic phase, the lattice is centrosymmetric and electric polarization is symmetry forbidden; however, antiferromagnetic ordering below $T_{\rm N}$ simultaneously breaks time-reversal symmetry and lowers the magnetic point-group symmetry, thereby allowing linear cross-coupling terms in the free energy of the form $
F_{\mathrm{ME}} = -\alpha_{ij}E_iH_j,$
which yield a field-induced polarization $P_i = \alpha_{ij}H_j$ and magnetization $M_i = \alpha_{ij}E_j$~\cite{dzyaloshinskii1960magneto,astrov1960magnetoelectric}. This symmetry-driven mechanism closely parallels that of the parent compound CNO, where antiferromagnetic order reduces the symmetry to C2/c$'$ and activates sizable off-diagonal ME tensor components, together with toroidal moments and magnetic-field-induced polarization~\cite{PhysRevB.93.075117}. Microscopically, several exchange pathways arising from structural distortion in the spin lattice convert spin correlations into electric dipoles: (i) inverse Dzyaloshinskii--Moriya interactions $\mathbf{P}\propto \mathbf{e}_{ij}\times(\mathbf{S}_i\times\mathbf{S}_j)$ associated with canting or noncollinearity~\cite{PhysRevLett.95.057205}, (ii) exchange striction $\mathbf{P}\propto(\mathbf{S}_i\cdot\mathbf{S}_j)$ that displaces oxygen ligands in collinear antiferromagnets~\cite{PhysRevB.73.094434}, and most importantly (iii) spin--orbit--assisted metal--ligand hybridization, where the strong orbital moment of Co$^{2+}$ directly modulates Co--O covalency, generating an electronic polarization~\cite{arima2007ferroelectricity}. Because these mechanisms scale with both magnetic susceptibility and spin-orbit coupling, the competing anisotropic exchanges and the enhanced magnetic frustration introduced by Zn substitution reduce the energy scale (i.e., $T_N$) separating nearly degenerate spin configurations, thereby enabling field-induced canting, domain redistribution, and reorientation of noncollinear spin correlations. This enhanced susceptibility of the spin configuration to external magnetic field strengthens the ME coefficient and, concomitantly, links dielectric, magnetic, and entropy responses consistent with the sizable MCE. Thus, CZNO extends the magnetically induced polarization mechanism established in CNO into a regime where spin-orbit coupling and competing exchange interaction promote both linear magnetoelectricity and multiferroicity within the same honeycomb framework.

The unavoidable lattice distortions in CZNO lower the local symmetry of the Co$^{2+}$ octahedra and modify the Co–O–Co exchange network, which—through strong spin–orbit coupling—induces pronounced magnetic anisotropy. In addition, Zn substitution perturbs the local structure and splits exchange pathways into inequivalent, competing interactions, leading to ground-state degeneracy and thereby resulting in a field-induced spin-flop–like metamagnetic transition (see inset of Fig.~\ref{Magnetocaloric}(c)) in the antiferromagnetic state of the distorted honeycomb lattice. A comparison of three energy scales for spin-flop transitions, long-range ordering, and the Curie-Weiss temperature for the A$_4$B$_2$O$_9$ family of materials is shown in Fig.~\ref{fig:PhaseDia}. The Curie-Weiss temperature, which is related to the exchange energy scale, is higher for the Nb$^{5+}$ variants than for their Ta$^{5+}$ counterparts. Although both Nb$^{5+}$ (4$d^0$) and Ta$^{5+}$ (5$d^0$) are formally $d^0$ ions, the larger spatial extent of the $5d$ orbitals and the stronger covalency associated with Ta can alter the orbital overlap along the superexchange pathways, thereby influencing the effective magnetic exchange. 
Furthermore, as we can see, the $\theta_\text{CW}$ is higher for the $A=\mathrm{Mn}^{2+}$ ($3d^5$) variant, while the $\mathrm{Co}^{2+}$ ($3d^7$) and $\mathrm{Fe}^{2+}$ ($3d^6$) variants are an order of magnitude lower than the $\mathrm{Mn}^{2+}$ compound in this series. The spin-flop field is weaker for Co- and Fe-rich compounds (see Fig.~\ref{fig:PhaseDia}).
This can be understood from the substantial reduction in the exchange scale, as reflected in the much lower Curie–Weiss temperatures compared to the $Mn^{2+}$ compound. Since the spin-flop field scales as $H_\text{C}\propto\sqrt{H_EH_A}$~\cite{PhysRevB.95.104418}, where $H_E$ and $H_A$ are the effective exchange and anisotropy fields, respectively, the reduction in the exchange field dominates over any enhancement in anisotropy arising from spin–orbit coupling, resulting in a lower critical field for spin reorientation. A small $H_\text{sf}$ can be advantageous for practical uses, especially for enabling low-power magnetic switching, and improving device functionality~\cite{mostovoy2024multiferroics}.
The effect of non-magnetic substitution in the A$_4$B$_2$O$_9$ series and the resulting evolution of the energy scales, as well as the magnetic frustration parameters, are summarized in Table~\ref{Table_3}. A systematic comparison within the niobate family shows how partial replacement by non-magnetic ions (Zn, Mg) progressively dilutes the magnetic sublattice, thereby weakening the dominant exchange interactions. Consequently the long-range magnetic ordering is suppressed, and the balance between competing interactions is altered, thereby determining whether the system stabilizes in an antiferromagnetic (AFM) or ferrimagnetic ground state. Such dilution due to controlled non-magnetic substitution enhances magnetic frustration and can drive the system toward reduced ordering temperatures or even destabilize conventional order altogether.

The strong field dependence of the magnetic entropy suggests that the low-energy spin manifold is highly compressible by magnetic fields~\cite{wolf2011magnetocaloric}. Thermodynamically, this implies the presence of nearly degenerate spin configurations separated by small energy scales, consistent with frustrated exchange interaction. In this context, the observation of MCE in this family of frustrated materials with distorted honeycomb network provides an independent probe of the same physics that underlies the ME and multiferroic responses—namely, a delicate balance of competing interactions that enhances susceptibility to external perturbations. Recently, synergistic efforts have been devoted towards identifying suitable frustrated materials exhibiting large MCE over a broad temperature range for practical applications. For instance, room-temperature operation is desirable for household refrigeration, whereas large MCE at low temperatures is essential for helium liquefaction and space technologies. 
From a broader perspective, the coexistence of moderate entropy changes along with magnetoelectric coupling suggests that the free energy landscape couples polarization not only to magnetization but also to entropy, unifying caloric and electric responses within a single framework.

\section{Conclusion}
In this work, we synthesized CZNO and investigated its crystal structure, magnetic susceptibility, and specific heat. The structural analysis from XRD data confirms the same trigonal space group, $P\bar{3}c1$, as its parent compound, CNO. The temperature-dependent magnetic susceptibility shows the AFM transition at $\sim 14$ K, and specific heat experiments also reveal a sharp lambda-like peak confirming the AFM transition owing to finite interlayer coupling between honeycomb planes. The large value of the Curie-Weiss temperature, $-70$ K, suggests the strong antiferromagnetic interactions between the Co$^{2+}$ spins, while an effective magnetic moment, 5.8 $\mu_B$ that is larger than that expected for the high spin ($S $= 3/2) value of Co$^{2+}$, supporting the presence of orbital angular momentum. The isothermal magnetization exhibits linear behavior up to 7 T, supporting strong antiferromagnetic interactions between Co$^{2+}$ moments. The antiferromagnetic transition temperature remains unaffected with a field of 9 T in specific heat, which suggests the robust antiferromagnetic interaction between the Co$^{2+}$ spins. The magnetic specific heat exhibits a nontrivial power-law, $C_m\sim T^{1.45}$ behavior below 14 K, suggesting the presence of unconventional low-energy spin excitations in this magnet. The MCE—possibly due to the combined effect of exchange frustration, anisotropic interactions, and field-induced spin reconfiguration in the host honeycomb lattice—is moderately small due to the strong antiferromagnetic exchange interactions between Co$^{2+}$ moments in the host spin lattice.

The presence of spin-orbit coupled Co$^{2+}$ moments in an octahedral environment places CZNO in the broader class of Kitaev magnets. The field-induced metamagnetic like state and the emergence of electric polarization below $T_\text{N}$ identify CZNO as a linear magnetoelectric antiferromagnet. The present material is a rare example of a type-II multiferroic, in which the magnetic order induces electric polarization. Moreover, the observed MCE effect at low temperatures, along with the field-induced magnetization and dielectric effects, suggests the cross-coupling of electric and magnetic order parameters, as well as spin-orbit entangled physics. Future studies exploring the effects of chemical pressure and high magnetic fields may provide deeper insights into the ground state of this family of honeycomb based frustrated antiferromagnet.

\section*{Acknowledgement}
P.K. acknowledges the funding by the Anusandhan National Research Foundation (ANRF), Department of Science and Technology, India through Research Grants.

\bibliography{CZNO}

\end{document}